\documentclass[fleqn,twoside]{article}
\usepackage{espcrc2}

\usepackage{epsf}                                    
\usepackage{graphics}                                
\usepackage{graphpap}   

\usepackage{graphicx}
\usepackage[figuresright]{rotating}

\newcommand{\AmS}{{\protect\the\textfont2
  A\kern-.1667em\lower.5ex\hbox{M}\kern-.125emS}}

\newcommand{\dzero}{\mbox{D\O}}

\newcommand{\et}{\mbox{$E_{T}$}}

\newcommand{\aeta}{\mbox{$|\eta|$}}				

\newcommand{\met}{\mbox{${\hbox{$E$\kern-0.63em\lower-.18ex\hbox{/}}}_{T}$}}
\newcommand{\metvec}{\mbox{${\hbox{$\vec{E}$\kern-0.63em\lower-.18ex\hbox{/}}}_{T}\,$}}
\newcommand{\metx}{\mbox{${\hbox{$E$\kern-0.63em\lower-.18ex\hbox{/}}}_{x}\,$}}
\newcommand{\mety}{\mbox{${\hbox{$E$\kern-0.63em\lower-.18ex\hbox{/}}}_{y}\,$}}

\def\D0{D\O}
\def\ETmiss{{\rm {\mbox{$E\kern-0.57em\raise0.19ex\hbox{/}_{T}$}}}}

\def\simge
{\mathrel{\rlap{\raise 0.53ex \hbox{$>$}}{\lower 0.53ex \hbox{$\sim$}}}}
\def\simle
{\mathrel{\rlap{\raise 0.53ex \hbox{$<$}}{\lower 0.53ex \hbox{$\sim$}}}}

\def\ETmiss{\mbox{${\hbox{$E$\kern-0.5em\lower-.1ex\hbox{/}\kern+0.15em}}_{\rm T}$}}

\def\1800{$\sqrt{s}=1800$ GeV}
\def\630{$\sqrt{s}=630$ GeV}

\title{The \dzero\ Detector for Run II}

\author{Levan Babukhadia%
	\address[SUNY]{State University of New York, Stony Brook, 
	New York 11794, U.S.A.}%
	\thanks{Current address: MS 357, Fermilab P.O. Box 500, Batavia, 
		Illinois, U.S.A. Electronic mail: blevan@fnal.gov.}
       (for the \dzero\ Collaboration)}
\begin{document}

\begin{abstract}
The general purpose \dzero\ collider detector at the Fermilab 
Tevatron has undergone major upgrades for Run II.  
We describe the current status and performance of the \dzero\ 
detector.
\vspace{1pc}
\end{abstract}

\maketitle

\setlength{\unitlength}{1.0mm}

\section{INTRODUCTION}

\begin{figure*}[!ht] \centering
  \begin{picture}(160,120)

  \put(-1,55){
    \begin{picture}(160,55)
      \includegraphics*[angle=0,width=8cm,height=6.6cm]{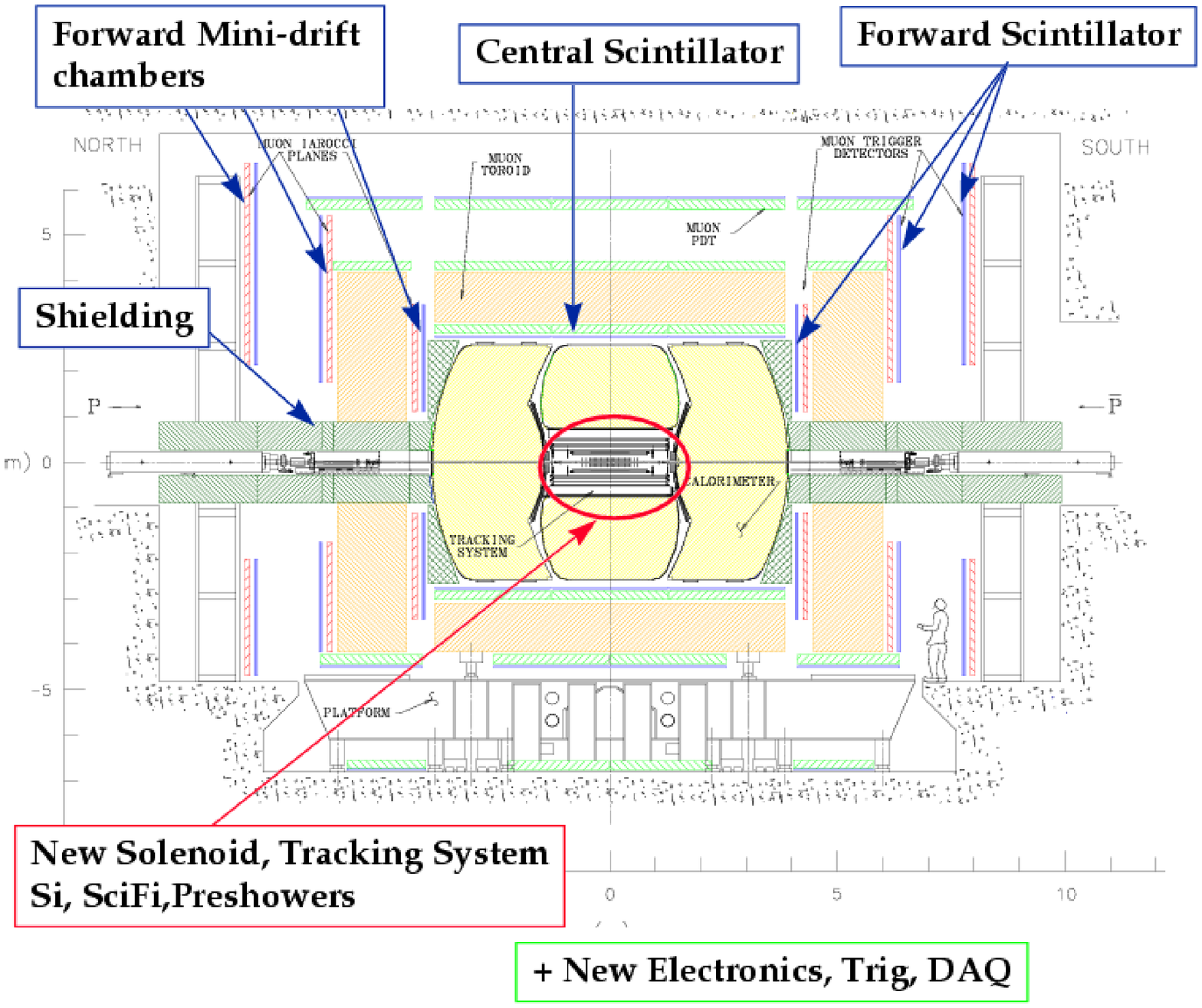}
    \end{picture}}

  \put(78,55){
    \begin{picture}(160,55)
      \includegraphics*[width=8.2cm,height=6.6cm]{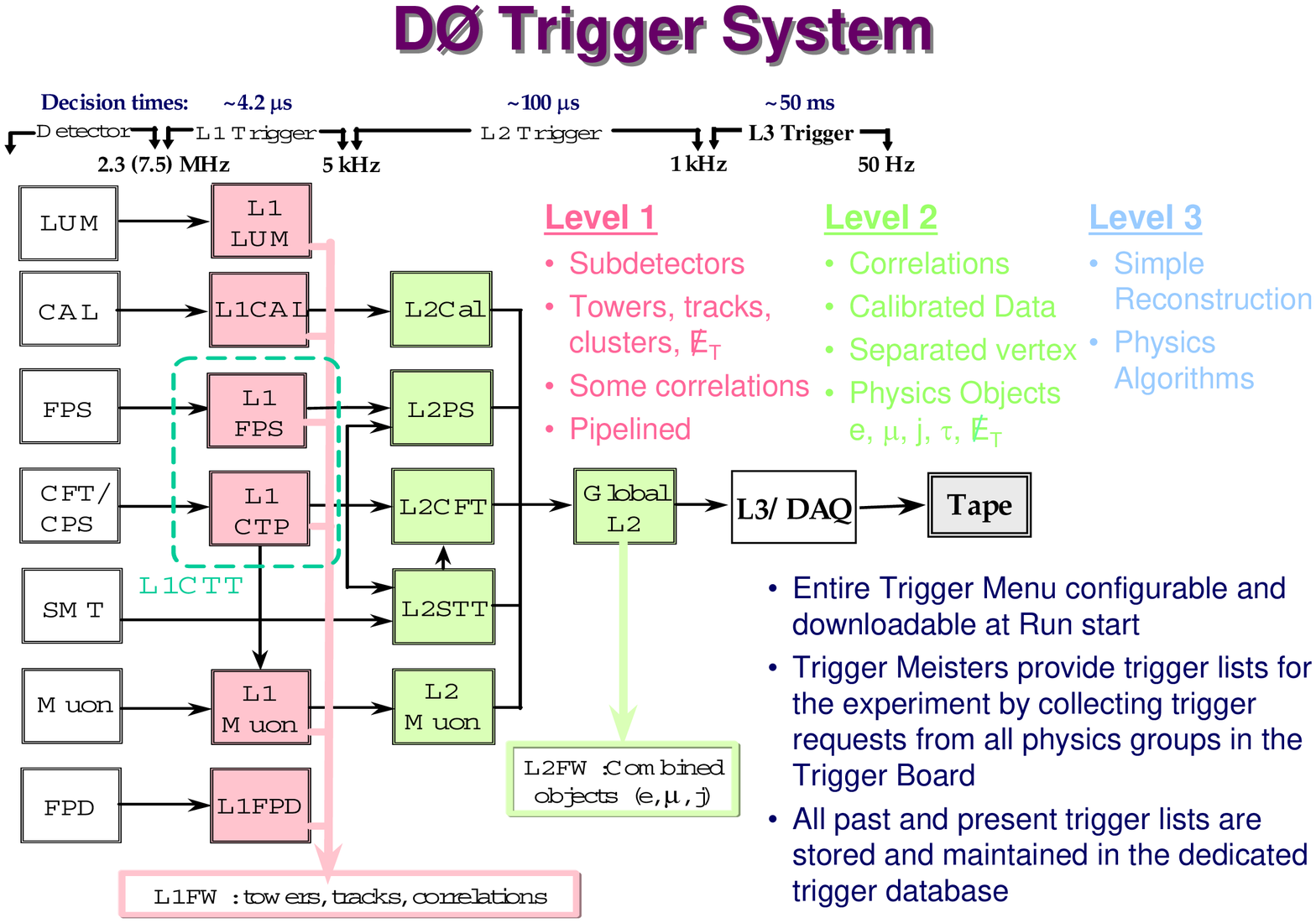}
    \end{picture}}

  \put(50,-5.5){
    \begin{picture}(53,55)
      \includegraphics*[width=6.3cm,height=6cm]{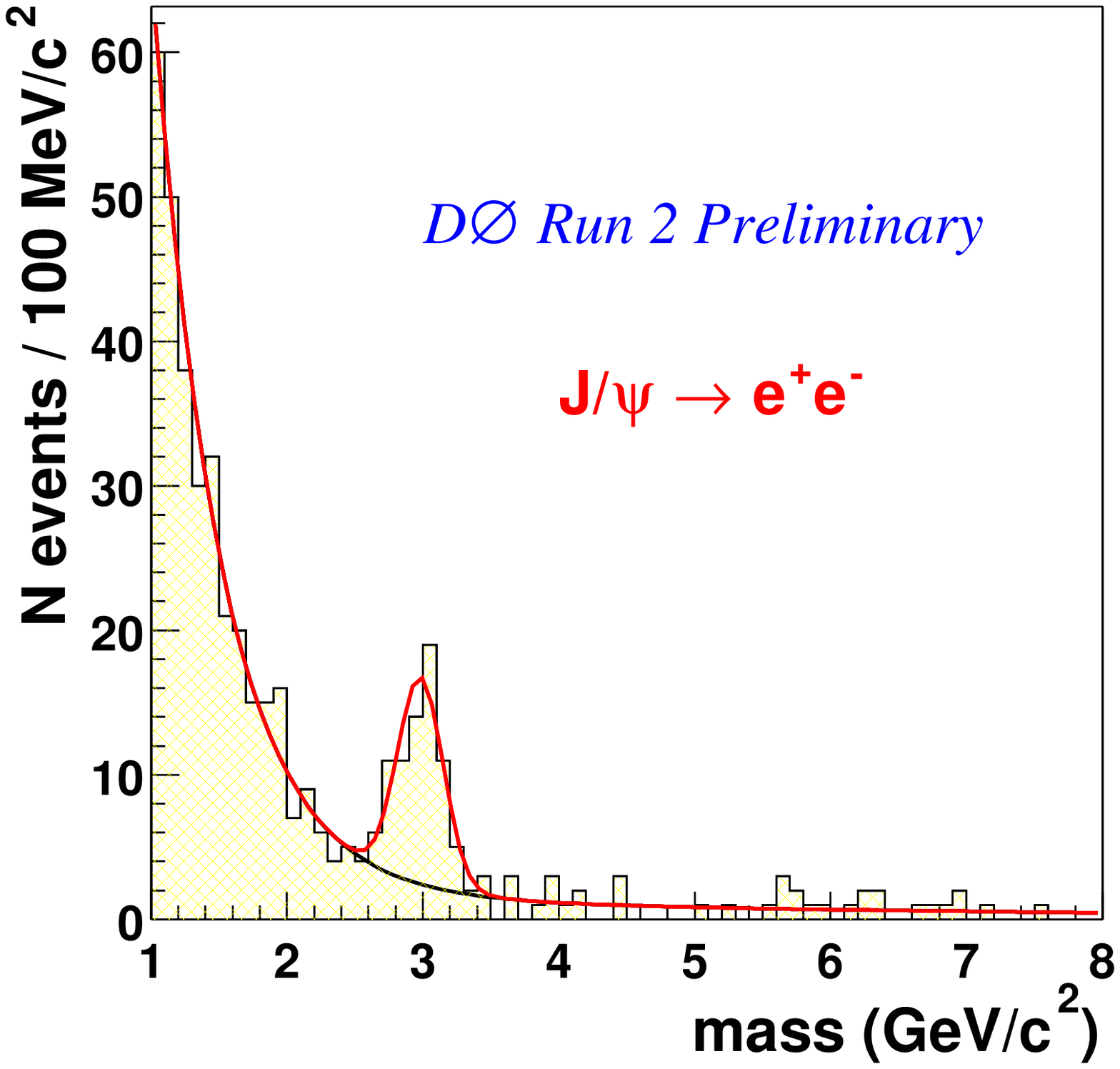}
    \end{picture}}

  \put(-1,2){
    \begin{picture}(53,50)
      \includegraphics*[width=5.3cm,height=5.0cm]{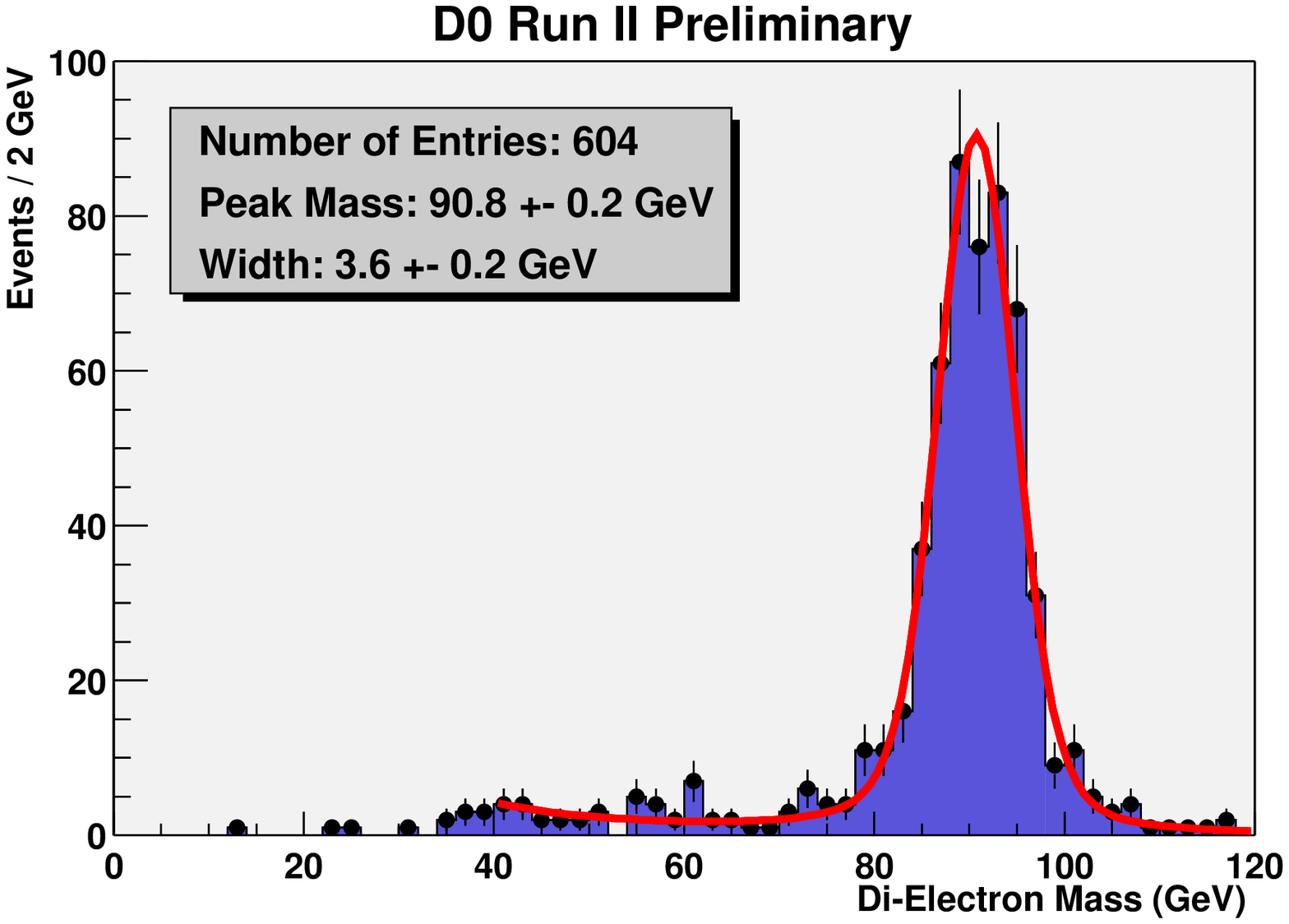}
    \end{picture}}

  \put(108,0){
    \begin{picture}(53,55)
      \includegraphics*[width=5.7cm,height=5.5cm]{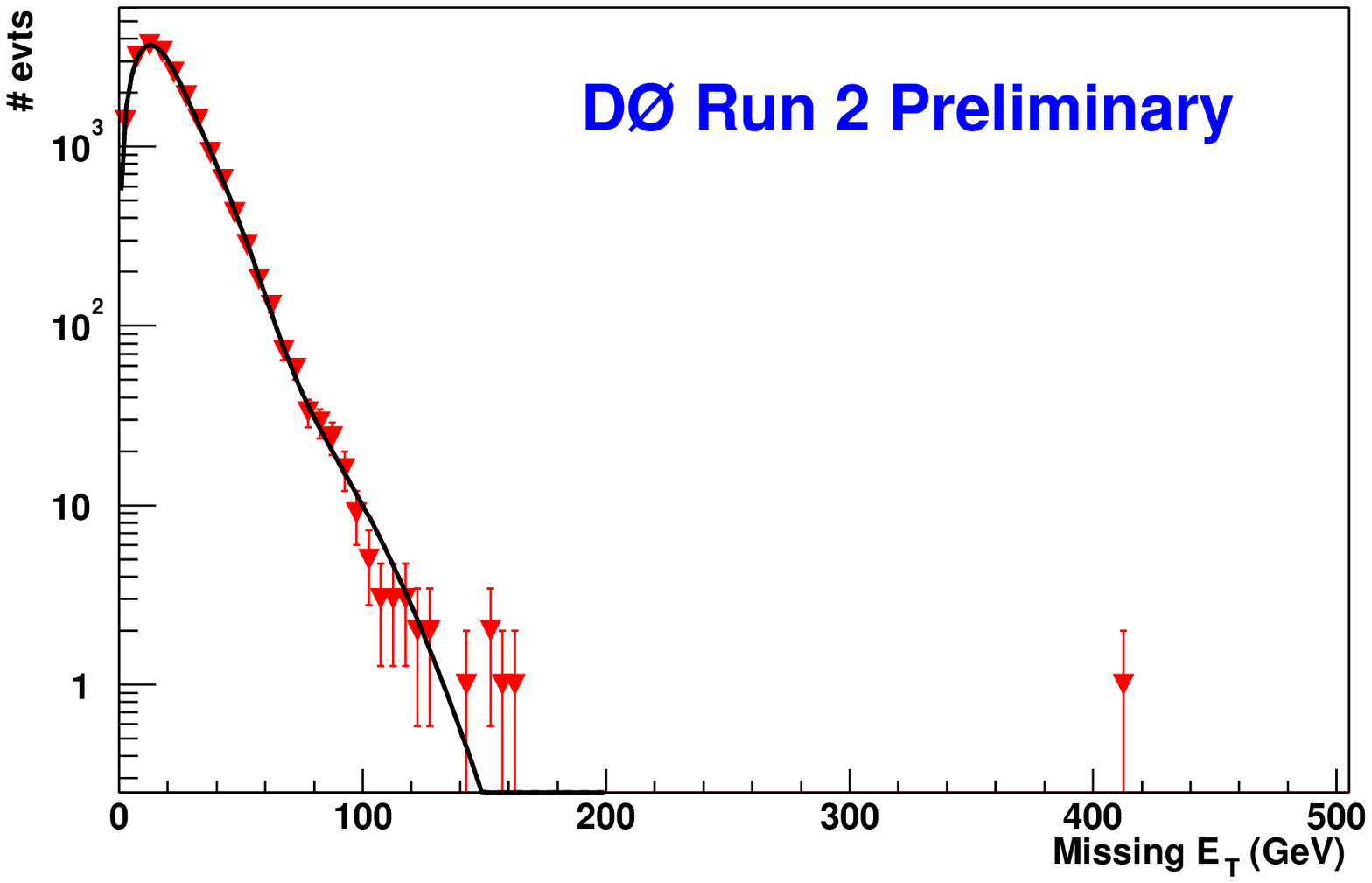}
    \end{picture}}

  \put(72,99){\makebox(0,0)[rb]{\scalebox{1.0}{ \bf (a) }}}       
  \put(159,85){\makebox(0,0)[rb]{\scalebox{1.0}{ \bf (b) }}}       
  \put(51.5,23){\makebox(0,0)[rb]{\scalebox{1.0}{ \bf (c) }}}       
  \put(107,23){\makebox(0,0)[rb]{\scalebox{1.0}{ \bf (d) }}}       
  \put(159,23){\makebox(0,0)[rb]{\scalebox{1.0}{ \bf (e) }}}       


  \end{picture}

  \vskip-0.7cm
  \caption{\footnotesize \dzero\ detector for Run II (a) and diagram 
           of the trigger system (b).  Invariant mass distribution of 
           di-EM objects showing $Z$-boson (c) and $J/\psi$ (d) 
           resonances.  Distribution of missing transverse energy 
           \met\ in multijet events (e).}

  \label{fig:fig1}
\end{figure*}

The Fermilab Tevatron has undergone significant upgrades for Run II.
Presently, it is delivering higher instantaneous luminosity than in 
Run I with a ten-fold decrease in the time between beam crossings.
The Run IIa will provide an integrated luminosity of 2 fb$^{-1}$,
about 20 times that of Run~I.
Also the center-of-mass energy of $p\bar{p}$ collisions has increased 
modestly, from 1.8 to 1.96 TeV. 

To achieve the physics goals for Run II which include searches for the 
Higgs boson, supersymmetry, extra dimensions, and other new phenomena, 
as well as precision studies of weak bosons, top quark, QCD, and 
$B$-physics, the collider detectors at the Tevatron need superb
electron, muon, and tau identification, excellent jet and missing
transverse energy (\met) reconstruction, and very good flavor tagging 
of jets through displaced vertices and leptons.

To meet these challenges in a high event rate environment, the \dzero\ 
detector has undergone major upgrades~\cite{d0upgrade}
(Fig.~\ref{fig:fig1}a).  
The central tracking system has been replaced by a silicon microstrip
tracker, surrounded by a scintillating fiber tracker, both 
immersed in a 2T magnetic field from a superconducting 
solenoid~\cite{ginther}.
To provide fast energy and position measurements for the electron 
trigger and offline electron identification, scintillating fiber
preshower detectors have been added in the central and forward regions.
Calorimeter trigger and readout electronics have been upgraded 
and the muon system has been significantly improved.
A new forward proton spectrometer (FPD) has been added for the 
diffractive and elastic physics program.
To handle high event rates, an entirely new trigger and data acquisition
system has been designed and implemented, see Fig.~\ref{fig:fig1}b.

\section{STATUS AND PERFORMANCE}

\begin{figure*}[!t] \centering
  \begin{picture}(160,105)

  \put(-1,54){
    \begin{picture}(53,50)
      \includegraphics*[width=5.3cm,height=5.0cm]{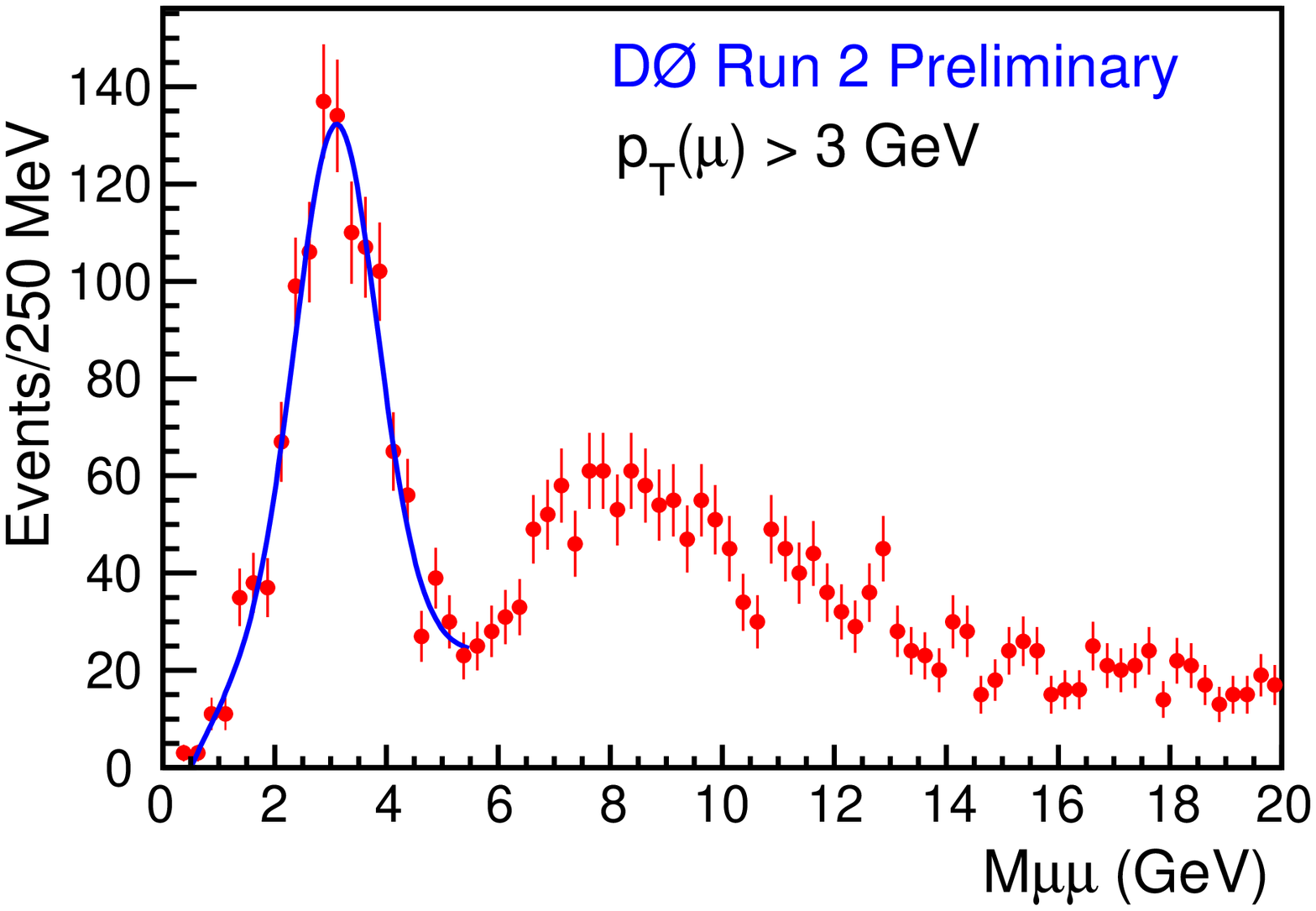}
    \end{picture}}

  \put(53,54){
    \begin{picture}(53,55)
      \includegraphics*[width=5.3cm,height=5.0cm]{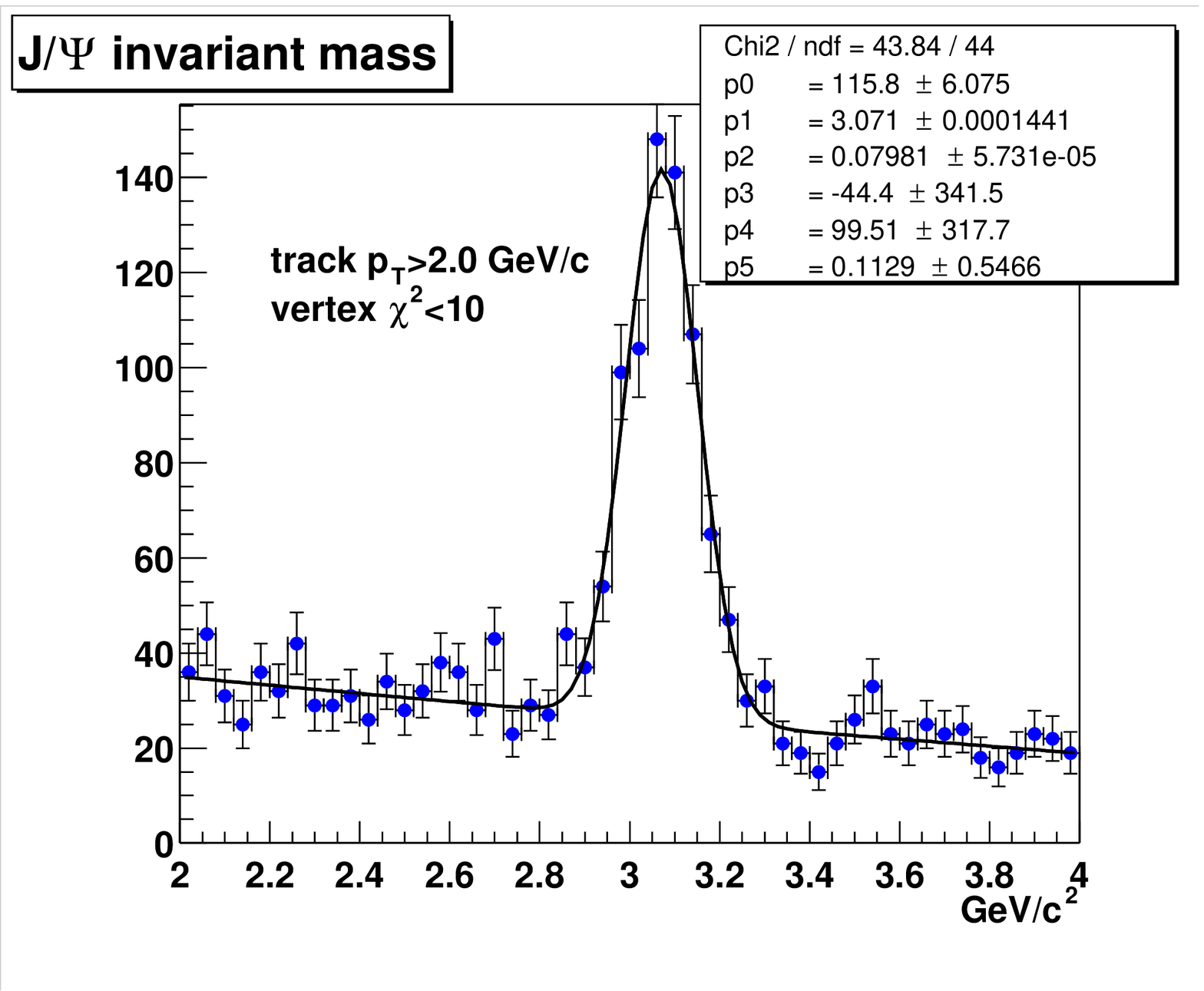}
    \end{picture}}

  \put(106,54){
    \begin{picture}(53,55)
      \includegraphics*[width=5.3cm,height=5.0cm]{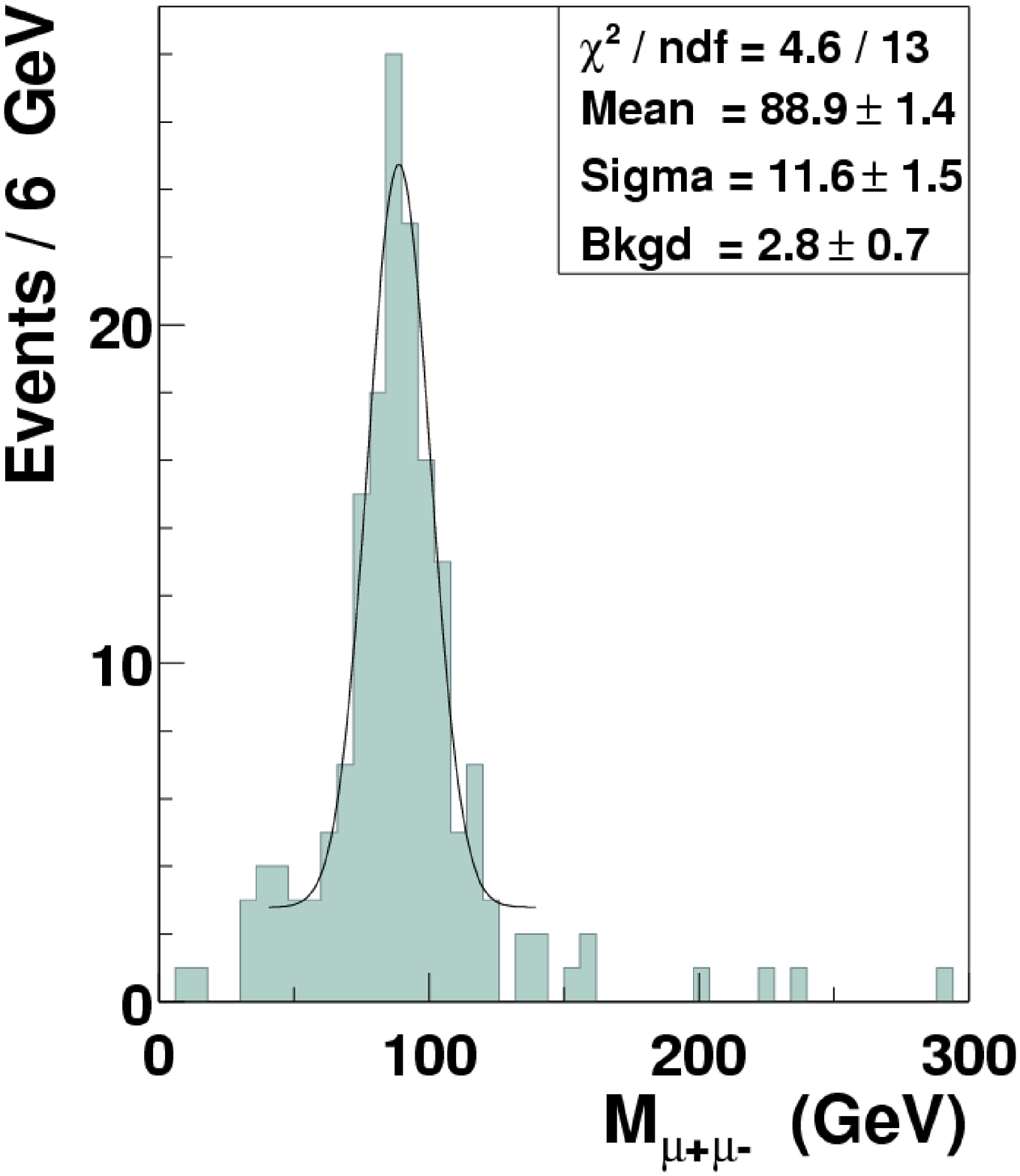}
    \end{picture}}

  \put(0,-1){
    \begin{picture}(160,55)
      \includegraphics*[width=16cm,height=5.5cm]{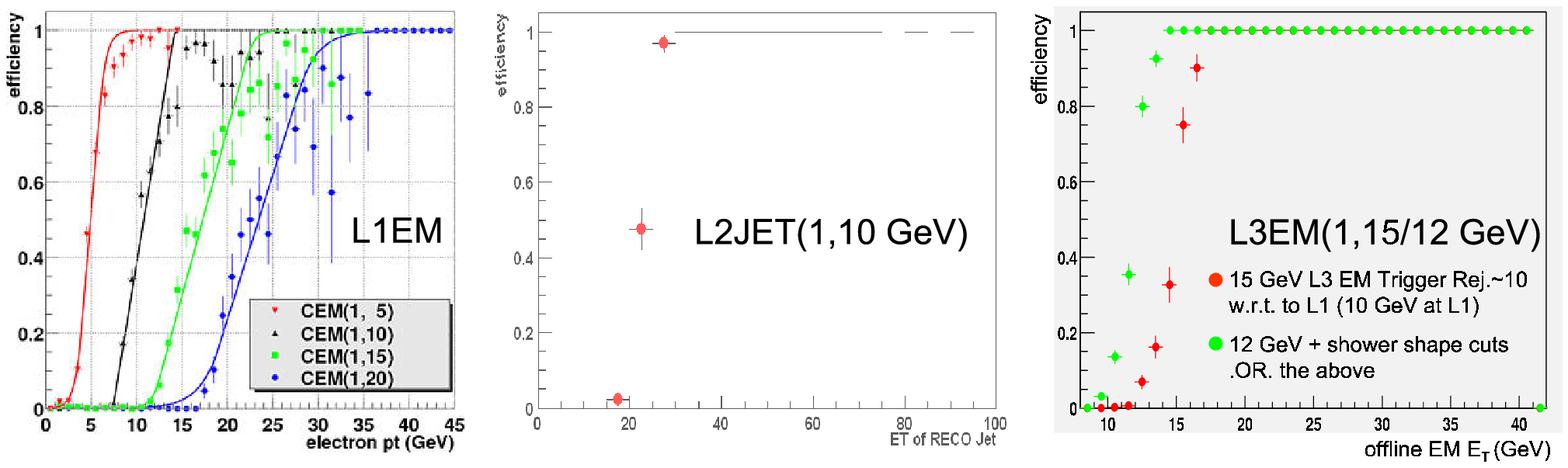}
    \end{picture}}

  \put(51.5,80){\makebox(0,0)[rb]{\scalebox{1.0}{ \bf (a) }}}       
  \put(101.5,80){\makebox(0,0)[rb]{\scalebox{1.0}{ \bf (b) }}}       
  \put(157,80){\makebox(0,0)[rb]{\scalebox{1.0}{ \bf (c) }}}       
  \put(48,33){\makebox(0,0)[rb]{\scalebox{1.0}{ \bf (d) }}}       
  \put(102,33){\makebox(0,0)[rb]{\scalebox{1.0}{ \bf (e) }}}       
  \put(157,33){\makebox(0,0)[rb]{\scalebox{1.0}{ \bf (f) }}}       


  \end{picture}

  \vskip-0.7cm
  \caption{\footnotesize Invariant mass distributions of di-muons using:
	   muon stand-alone system showing $J/\psi$ resonance (a) and 
	   muons matched with central tracks showing $J/\psi$ (b) and 
	   $Z$-boson (c) resonances.
	   Efficiencies for triggering on electromagnetic and hadronic
	   objects of different \et\ thresholds in the first (d), second 
           (e), and third (f) stages of the \dzero\ trigger system.}

  \label{fig:fig2}
\end{figure*}

{\bf Calorimeters.}  The nearly compensating, uniform, and 
hermetic liquid-argon uranium calorimeters from Run I, with 
completely upgraded readout and trigger electronics, provide 
about 50k readout cells of electromagnetic (EM) and fine and 
coarse hadronic compartments with less than 0.1\% dead or 
noisy channels.
About 5k calorimeter towers are also used for fast trigger 
decisions.
The system is fully commissioned.

For an \emph{in situ} calibration of EM calorimeters the 
$Z\rightarrow ee$ signal is employed.
Fig.~\ref{fig:fig1}c shows the $Z$ peak in a di-EM invariant 
mass distribution with close to expected resolution.
On the other end of the spectrum, Fig.~\ref{fig:fig1}d shows 
the $J/\psi$ resonance.
The calorimeter response to jets is calibrated using photon-jet 
data.
Fig.~\ref{fig:fig1}e shows the distribution of \met\ in 
multijet events, spanning several orders of magnitude; the 
present \met\ resolution is about 7 GeV.

{\bf The Muon system} consists of central and forward regions 
providing coverage up to $\aeta=2$ with three layers of 
scintillators and drift tubes, one inside and two outside the 
magnetized iron toroids.
In the central region, there are over 6k cells of proportional
drift tubes and 630 (360) scintillating counters inside 
(outside) the toroid.
In the forward regions, there are about 6k 8-cell mini drift 
tubes in eight octants per layer and over 4k forward 
scintillation counters or pixels.
The muon system is fully commissioned.

Fig.~\ref{fig:fig2}a shows the invariant mass distribution of 
di-muons from the muon stand-alone system, with the $J/\psi$ 
resonance visible above background.
Matching muons with central tracks significantly improves muon
transverse momentum resolution and Figs.~\ref{fig:fig2}b 
and~\ref{fig:fig2}c show the $J/\psi$ and the $Z$-boson 
resonances, respectively.
Timing cuts on muons reduce backgrounds from cosmic rays, and 
could aid in detection of slow moving particles.

{\bf The FPD system} consists of two arms of 18 Roman pots in 
four quadrupole and two dipole ``castles.''
Hits in scintillating fiber detectors installed in the Roman 
pots will be used for measuring fractional energy lost by the 
proton and scattering angle, and for triggering on elastic, 
diffractive, and double pomeron events.
Currently the system is in commissioning.
Several million elastic events have been recorded by FPD 
stand-alone system, and integration within \dzero\ is ongoing.

{\bf The \dzero\ Trigger} is a three stage hardware and 
software system that reduces the raw event rate of about 7 MHz 
down to under 50 Hz for recording events to tape
(Fig.~\ref{fig:fig1}b).
The hardware triggers in Level 1 (L1) allow triggering on EM and 
hadronic calorimeter objects, tracks, and muons. 
In L2 more physics-like objects are used employing correlations 
and allowing triggering on separated vertices.
In L3, physics algorithms are employed after fast event 
reconstruction.

Fig.~\ref{fig:fig2}d shows the efficiency ``turn-on'' for triggering
on EM objects with various transverse energy (\et) thresholds 
using the L1 calorimeter trigger.
Similarly, Fig.~\ref{fig:fig2}e shows the efficiency for triggering 
on a L2 ``jet'' of $\et=10$ GeV.
Finally, Fig.~\ref{fig:fig2}f shows the performance of L3 software
filter to trigger on $\et=15$ GeV EM objects, also indicating that 
the \et\ threshold can effectively be lowered by utilizing EM shower 
shape requirements.

\section{SUMMARY AND OUTLOOK}

The \dzero\ detector for Run II is operating and collecting physics 
data.
There has been enormous progress over the past year in installation, 
integration, and commissioning of the detector, and understanding the 
data.
Performance of the Run~II \dzero\ detector is very encouraging, all 
subdetectors are operating well, software and computing systems are 
working well, we are reconstructing electrons, muons, jets, \met, 
$J/\psi$'s, $W$'s and $Z$'s.
We are working hard to complete commissioning, improve calibration 
and alignment, optimize detector, trigger, and DAQ performance, and 
are on the way to exciting physics, with first physics results 
now appearing~\cite{narain}.


\end{document}